\documentclass[11pt,fleqn,twoside]{article}
\usepackage{amsfonts,amssymb,latexsym}
\makeatletter
\newcommand{\prava}[1]{\small\it
\begin{flushleft}
Copyright \copyright \ 1999 by  #1
\end{flushleft}}

\newcommand{\name}[1]{\begin{flushleft}
                       \LARGE \bf #1
                       \end{flushleft}\vspace{-3mm}}

\newcommand{\Author}[1]{\begin{flushleft}
                       \it #1 \end{flushleft}}

\newcommand{\Adress}[1]{\begin{flushleft}
                       \it #1 \end{flushleft}}

\newcommand{\Date}[1]{\begin{flushleft}
                      \small  \it #1 \end{flushleft}}

\newcommand{\ehkol}{Author \ name}
\newcommand{\ohkol}{Article \ name}
\renewcommand{\@evenhead}{
\hspace*{-3pt}\raisebox{-15pt}[\headheight][0pt]{\vbox{\hbox to \textwidth 
{\thepage \hfil \ehkol}\vskip4pt \hrule}}}
\renewcommand{\@oddhead}{
\hspace*{-3pt}\raisebox{-15pt}[\headheight][0pt]{\vbox{\hbox to \textwidth 
{\ohkol \hfil \thepage}\vskip4pt\hrule}}}
\renewcommand{\@evenfoot}{}
\renewcommand{\@oddfoot}{}

\newcommand{\be}{\begin{equation}}
\newcommand{\ee}{\end{equation}}
\newcommand{\ba}{\hspace*{-5pt}\begin{array}}
\newcommand{\ea}{\end{array}}
\newcommand{\p}{\partial}
\newcommand{\ds}{\displaystyle}
\makeatother

\begin{document}
\setcounter{page}{5}
\renewcommand{\theequation}{\arabic{section}.\arabic{equation}}

\thispagestyle{empty}

\renewcommand{\ehkol}{Z. Jiang}
\renewcommand{\ohkol}{Neumann and Bargmann Systems and
the Coupled KdV Hierarchy}

\begin{flushleft}
\footnotesize \sf
Journal of Nonlinear Mathematical Physics \qquad 1999, V.6, N~1,
\pageref{jiang-fp}--\pageref{jiang-lp}.
\hfill {\sc Letter}
\end{flushleft}

\vspace{-5mm}

\renewcommand{\footnoterule}{} {\renewcommand{\thefootnote}{}
\footnote{\prava{Z. Jiang}}}

\name{Neumann and Bargmann Systems Associated with an Extension of the
Coupled KdV Hierarchy} \label{jiang-fp}

\Author{Zhimin JIANG}

\Adress{Department of Mathematics, Shangqiu Teachers College,
Shangqiu 476000, China}

\Date{Received October 16, 1998;  Accepted December 03, 1998}

\begin{abstract}
\noindent
An eigenvalue problem with a reference function and the corresponding
hierarchy of nonlinear evolution equations are proposed. The
bi-Hamiltonian structure of the hierarchy is established by using the
trace identity. The isospectral problem is nonlinearized as to be
f\/inite-dimensional completely integrable systems in Liouville sense
under Neumann and Bargmann constraints.
\end{abstract}

\section{Introduction}

A major dif\/f\/iculty in theory of integrable systems is that there
is to date no completely systematic method for choosing properly an
isospectral problem $\psi_{x}=M\psi$ so that the zero-curvature
representation $M_{t}-\overline{N}_{x}+[M,\overline{N}]=0$ is nontrivial. By
inserting a reference function into AKNS and WKI isospectral
problems, we have obtained successfully two new hierarchies~[1, 2].

The coupled KdV hierarchy associated with the isospectral problem
\begin{equation}
\psi_{x}=M\psi,
\qquad M=\left(\begin{array}{cc}
\ds -\frac{1}{2}\lambda+\frac{1}{2}u&-v
\vspace{3mm}\\
1&\ds \frac{1}{2}\lambda-\frac{1}{2}u \end{array} \right)
\end{equation}
is discussed by D.~Levi, A.~Sym and S.~Wojciechowsk~[3]. The isospectral
problem~(1.1) has been nonlinearized as f\/inite-dimensional
completely integrable systems in Liouville sense~[4].

In this paper, we introduce the eigenvalue problem
\begin{equation}
\psi_{x}=M\psi,\qquad M=\left(\begin{array}{cc}
\ds -\frac{1}{2}\lambda+\frac{1}{2}u&-v
\vspace{3mm}\\
f(v)&\ds \frac{1}{2}\lambda-\frac{1}{2}u \end{array} \right),
\end{equation}
where $u$ and $v$ are two scalar potentials, $\lambda$ is a constant
spectral parameter and $f(v)$ called reference function is an
arbitrary smooth function.  The bi-Hamiltonian structure of the
corresponding hierarchy is established by using the trace
identity~[5, 6]. Since the reference function $f(v)$ in~(1.2) can be
chosen arbitrarily, many new hierarchies and their Hamiltonian forms
are obtained.When $f=(-v)^{\beta}$ $(\beta\geq0)$, the isospectral
problem (1.2) is nonlinearized as f\/inite-dimensional completely
integrable sysstems in Liouville sense under Neumann and Bargmann
constraints between the potentials and eigenfunctions.

\setcounter{equation}{0}
\section{Preliminaries}

Consider the adjoint representation of (1.2)
\begin{equation}
N_{x}=MN-NM,\qquad 
N=\left( \begin{array}{cc} a&b\\ c&-a \end{array}
\right)=\sum\limits_{j=0}^{\infty}\left(
\begin{array}{cc} a_i&b_i\\ c_i&-a_i \end{array} \right)
\lambda^{-j}
\end{equation}
which leads to
\begin{equation}
c_0=b_0=0, \qquad a_0=-\frac{1}{2}\alpha \quad ({\rm constant}),
\end{equation}
\begin{equation}
c_1={\alpha}f(v),\qquad b_1=-{\alpha}v,\qquad a_1=0,
\end{equation}
\begin{equation}
c_2={\alpha}(f^{\prime}(v)v_x+uf(v)),
\qquad b_2={\alpha}(v_x-uv),
\qquad a_2=-{\alpha}vf(v),
\end{equation}
\begin{equation}
a_j=-\partial^{-1}(vc_j+f(v)b_j),
\end{equation}
\begin{equation}
\left( \begin{array}{c} c_1\\ b_1 \end{array} \right)=\alpha\left(
\begin{array}{c}
f(v)\\ -v \end{array} \right), 
\qquad
\left( \begin{array}{c} c_{j+1}\\
b_{j+1} \end{array} \right)=L\left( \begin{array}{c} c_j\\ b_j
\end{array}
\right), \qquad j=1,2,\ldots,
\end{equation}
where
$\ds \partial=\frac{d}{dx}$, $\partial\partial^{-1}=\partial^{-1}\partial=1$,
\[
L=\left( \begin{array}{cc}
{\partial+u+2f\partial^{-1}v}&{2f\partial^{-1}f}
\vspace{1mm}\\
{-2v\partial^{-1}v}&{-\partial+u-2v\partial^{-1}f} \end{array}
\right).
\]

It is easy from (1.2) and (2.1) to calculate that
\[
{\rm tr}\left(N\frac{\partial{M}}{\partial\lambda}\right)=-a, 
\qquad {\rm tr}\left(N\frac{\partial{M}}{\partial{u}}\right)=a, 
\qquad {\rm
tr}\left(N\frac{\partial{M}}{\partial{v}}\right)=-c+f^{\prime}(v)b. 
\]

Noticing the trace identity~[5, 6]
\[
\left(\frac{\delta}{\delta{u}},\frac{\delta}{\delta{v}}\right)(-a)=
\frac{\partial}{\partial\lambda}(a,-c+f^{\prime}(v)b),
\]
hence we deduce that
\begin{equation}
\left(\frac{\delta}{\delta{u}},\frac{\delta}{\delta{v}}\right)H_j
=\left(G^{(1)}_{j-2},G^{(2)}_{j-2}\right), 
\qquad H=\frac{a_{j+1}}{j},
\end{equation}
where
\begin{equation}
G^{(1)}_{j-2}=a_j, \qquad G^{(2)}_{j-2}=-c_j+f^{\prime}(v)b_j.
\end{equation}

\setcounter{equation}{0}
\section{The hierarchy and its Hamiltonian structure}

Let $\psi$ satisfy the isospectral problem (1.2) and the auxiliary
problem
\begin{equation}
\psi_{t}=\overline{N}\psi,\qquad \overline{N} =\left( \begin{array}{cc}
A&B\\ C&-A \end{array} \right),
\end{equation}
where 
\[
A=A_m+\sum\limits_{j=0}^{m-1}a_{j}\lambda^{m-j},
\qquad B=\sum\limits_{j=1}^{m}b_{j}\lambda^{m-j},
\qquad C=\sum\limits_{j=1}^{m}c_{j}\lambda^{m-j}.
\]
 The compatible condition
$\psi_{xt}=\psi_{tx}$ between (1.1) and (3.1) gives the
zero-curvature representation $M_t-\overline{N}_x
+[M,\overline{N}]=0$, from which we have
\[
A_{m}=w(\partial+u)c_m+wf^{\prime}(v)(\partial-u)b_m,
\]
\begin{equation}
\left( \begin{array}{c} u_t\\ v_t \end{array} \right)=\theta_0L\left(
\begin{array}{c}
c_m\\ b_m \end{array} \right)=\theta_0\left( \begin{array}{c}
c_{m+1}\\ b_{m+1} \end{array} \right),
\end{equation}
where $\ds w=\frac{1}{2}(vf^{\prime}(v)+f)^{-1}$,
\begin{equation}
\theta_{0}=\left( \begin{array}{cc}
2\partial{w}&-2\partial{w}f^{\prime}(v)\\ 2wv&2wf \end{array}
\right).
\end{equation}
By (2.6) we know that Eqs.(3.2) are equivalent to the hierarchy of
nonlinear evolution equations
\begin{equation}
\left( \begin{array}{c} u_{t}\\ v_{t} \end{array}
\right)=\theta_{0}L^{m}\left(
\begin{array}{c} \alpha{f(v)}\\ -\alpha{v} \end{array}
\right),\qquad m=1,2,\ldots .
\end{equation}
Let the potentials $u$ and $v$ in (1.2) belong to the Schwartz space
$S(-\infty,+\infty)$ over $(-\infty,+\infty)$. Noticing (2.5) and
(2.8) we get
\begin{equation}
\left( \begin{array}{c} c_{j}\\ b_{j} \end{array}
\right)=\theta_{1}\left(
\begin{array}{c} G^{(1)}_{j-2}\vspace{2mm}
\\ G^{(2)}_{j-2} \end{array} \right),
\qquad
\theta_{1}=\left(
\begin{array}{cc} {-2wf^{\prime}(v)\partial}&{-2wf}\\
{-2w\partial}&{2wv} \end{array} \right).
\end{equation}

Then the recursion relations (2.5), (2.6) and the hierarchy (3.2) can
be written as
\[
G_{-2}=-\frac{1}{2}\alpha(1,0)^{T}, \quad
G_{-1}=-\alpha(0,vf^{\prime}(v)+f)^{T}, \quad 
G_{0}=-\alpha(vf,uf+uvf^{\prime}(v))^{T},
\]
\begin{equation}
KG_{j-1}=JG_{j},
\end{equation}
\begin{equation}
(u_t,v_t)^{T}=JG_{m-1}=KG_{m-2},
\end{equation}
where $J=\theta_{0}\theta_{1}$ and $K=\theta_{0}L\theta_{1}$ are two
skew-symmetric operators,
\[
J=\left( \begin{array}{cc} 0&-2\partial{w}\\ -2w\partial&0
\end{array} \right),
\qquad 
K=\left( \begin{array}{cc} K_{11}&K_{12}\\ K_{21}&K_{22}
\end{array}
\right),
\]
in which
\[
\left\{
\begin{array}{ll}
K_{11}=&-2\partial-4\partial{w}(\partial{f^{\prime}(v)
+f^{\prime}(v)\partial})w\partial,\vspace{1mm}\\
K_{12}=&-2\partial{wu}+4\partial{w}(f^{\prime}(v)\partial{v}-\partial{f})w,
\vspace{1mm}\\
K_{21}=&-2wu\partial+4w(f\partial-v\partial{f^{\prime}(v)})w\partial,
\vspace{1mm}\\
K_{22}=&-4w(v\partial{f}+f\partial{v})w.  \end{array}
\right.
\]

From (2.7) we obtain the desired bi-Hamiltonian form of (3.7)
\begin{equation}
\left( \begin{array}{c} u_{t}\\ v_{t} \end{array} \right)=J\left(
\begin{array}{c}
\ds \frac{\delta}{\delta{u}}\vspace{2mm}\\ 
\ds \frac{\delta}{\delta{v}} \end{array}
\right)H_{m+1}=K\left(
\begin{array}{c} \ds \frac{\delta}{\delta{u}}\vspace{2mm}\\ 
\ds \frac{\delta}{\delta{v}}
\end{array}
\right)H_{m}.
\end{equation}

\setcounter{equation}{0}
\section{Nonlinearization of the isospectral problem}

Let $\lambda_{j}$ and $\psi(x)=(q_{j}(x),p_{j}(x))^{T}$ be eigenvalue
and the associated eigenfunction of (1.2). Through direct
verif\/ication we know that the functional gradient
$\ds \nabla_{(u,v)}\lambda_{j}
=\left(\frac{\delta\lambda_j}{\delta{u}},
\frac{\delta\lambda_j}{\delta{v}}\right)$
satisf\/ies
\begin{equation}
\nabla_{(u,v)}\lambda_{j}=\left(q_jp_j,-p_j^2-f^{\prime}(v)q_j^2\right),
\end{equation}
\begin{equation}
\theta_{1}\nabla\lambda_{j}=\left( \begin{array}{c} p_{j}^{2}\\
-q_{j}^{2} \end{array} \right), 
\qquad L\left( \begin{array}{c} p_{j}^{2}\\
-q_{j}^{2} \end{array} \right)=\lambda_{j}\left( \begin{array}{c}
p_{j}^{2}\\ -q_{j}^{2} \end{array} \right)
\end{equation}
in view of (1.2). Substituting the f\/irst expression of (4.2) into
the second expression and acting with $\theta_{0}$ upon once, we have
\begin{equation}
K\nabla\lambda_{j}=\lambda_{j}J\nabla\lambda_{j}.
\end{equation}
So, the Lenard operator pair $K$, $J$ and their gradient series
$G_{j}$ satisfy the basic conditions (3.6) and (4.3) given in
Refs.~[7, 8] for the nonlinearization of the eigenvalue problem~(1.2).

\medskip

\noindent
{\bf Proposition 4.1.} {\it When $f(v)=(-v)^{\beta}$ $(\beta\geq0)$,
the isospectral problem $(1.2)$ 
can be nonlinearized as to be a Neumann system. }
\medskip

In fact, the Neumann constraint
$G_{-1}|_{\alpha=1}=\sum\limits_{j=1}^{N} \nabla\lambda_{j}$ gives
\begin{equation}
\langle{q},p\rangle=0,
\langle{p},p\rangle=
(\beta+1)(-v)^{\beta}+\beta(-1)^{\beta-1}\langle{q},q\rangle.
\end{equation}
By dif\/ferentiating (4.4) with respect to $x$ and using (1.2), we
have
\begin{equation}
\left\{
\begin{array}{l}
\ds u=\frac{1}{\beta+1}\left(\frac{\langle\Lambda{p},p\rangle}
{\langle{p},p\rangle}+ \beta\frac{\langle\Lambda{q},
q\rangle}{\langle{q},q\rangle}\right),\vspace{3mm}\\
v=\langle{q},q\rangle.  \end{array}
\right.
\end{equation}
Substituting (4.5) into the equations for the eigenfunctions
\begin{equation}
\left(\begin{array}{c}
q_{jx}\\p_{jx}
\end{array}\right)
=\left(
\begin{array}{cc}
\ds -\frac{1}{2}\lambda_{j}+\frac{1}{2}u&-v
\vspace{2mm}\\
(-v)^{\beta}&\ds \frac{1}{2}\lambda_{j}-\frac{1}{2}u\end{array}
\right)
\left(\begin{array}{c}q_j\\p_j\end{array}\right),
\qquad j=1,\ldots,N,
\end{equation}
we obtain the Neumann system
\begin{equation}
\left\{ \begin{array}{l}
\ds q_{x}=-\frac{1}{2}\Lambda{q}-\langle{q},q\rangle{p}+
\frac{1}{2(\beta+1)}
\left(\frac{\langle\Lambda{p},p\rangle}{\langle{p},p\rangle}+
\beta\frac{\langle\Lambda{q},q\rangle}{\langle{q},q\rangle}\right)q,
\vspace{3mm}\\
\ds p_{x}=\frac{1}{2}\Lambda{p}+\langle{p},p\rangle{q}-
\frac{1}{2(\beta+1)}
\left(\frac{\langle\Lambda{p},p\rangle}{\langle{p},p\rangle}+ 
\beta\frac{\langle\Lambda{q},q\rangle}{\langle{q},q\rangle}\right)p,
\vspace{3mm}\\
\langle{p},p\rangle=(-1)^{\beta}\langle{q},q\rangle^{\beta},
\qquad \langle{q},p\rangle=0.
\end{array} \right.
\end{equation}
where $p=(p_1,\ldots,p_N)^T$, $q=(q_1,\ldots,q_N)^T$, $\Lambda={\rm
diag}(\lambda_1, \ldots,\lambda_N)$, and $\langle,\rangle$ stands for
the canonical inner product in ${\bf R}^{N}$.

\medskip

\noindent
{\bf Proposition 4.2.} {\it When $f(v)=(-v)^{\beta}$ $(\beta\geq0)$,
the isospectral problem $(1.2)$ can be nonlinearized as to be a
Bargmann system.  }

\medskip

In fact, the Bargmann constraint
$G_{0}|_{\alpha=1}=\sum\limits_{j=1}^{N} \nabla\lambda_{j}$ gives
\begin{equation}
\left\{
\begin{array}{l}
\ds u=\frac{1}{\beta+1}\langle{p},p\rangle\langle{q},p\rangle
^{-\frac{\beta}{\beta+1}}-\frac{\beta}{\beta+1}\langle{q},q\rangle
\langle{q},p\rangle^{-\frac{1}{\beta+1}},
\vspace{3mm}\\
\ds v=-\langle{q},p\rangle^{\frac{1}{\beta+1}}.  \end{array}
\right.
\end{equation}
Substituting (4.8) into (4.6), we obtain the f\/inite-dimensional
Hamiltonian system
\begin{equation}
\left\{
\begin{array}{l}
\ds q_x=-\frac{1}{2}\Lambda{q}+\langle{q},p\rangle^{\frac{1}{\beta+1}}p
+\frac{1}{2(\beta+1)}\langle{p},p\rangle\langle{q},p\rangle
^{-\frac{\beta}{\beta+1}}q
\vspace{3mm}\\
\ds \qquad \qquad -\frac{\beta}{2(\beta+1)}\langle{q},q\rangle
\langle{q},p\rangle^{-\frac{1}{\beta+1}}q=\frac{\partial{H}}{\partial{p}},
\vspace{3mm}\\
\ds p_x=\frac{1}{2}\Lambda{p}-\frac{1}{2(\beta+1)}\langle{p},
p\rangle\langle{q}, p\rangle^{-\frac{\beta}{\beta+1}}p+
\langle{q},p\rangle^{\frac{\beta}{\beta+1}}
\vspace{3mm}\\
\ds \qquad \qquad
+\frac{\beta}{2(\beta+1)}\langle{q},q\rangle\langle{q},p\rangle 
^{-\frac{1}{\beta+1}}p=-\frac{\partial{H}}{\partial{q}}.  
\end{array} \right.
\end{equation}
The Hamiltonian is
\[ 
H=-\frac{1}{2}\langle{\Lambda}q,p\rangle
+\frac{1}{2}{\langle}p,p{\rangle}{\langle}q,p{\rangle}^{\frac{1}{\beta+1}}
-\frac{1}{2}{\langle}q,q{\rangle}{\langle}q,p{\rangle}^{\frac{\beta}
{\beta+1}}. 
\]

\setcounter{equation}{0}
\section{Integrability of the Neumann system}

The Poisson brackets of two functions in symplectic space $({\bf
R}^{2N},dp\wedge{dq})$ are def\/ined as
\[
(F,G)=\sum\limits_{j=1}^{N}\left(\frac{\partial{F}}{\partial{q_j}}
\frac{\partial{G}}{\partial{p_j}}
-\frac{\partial{F}}{\partial{p_j}}
\frac{\partial{G}}{\partial{q_j}}\right)=
\langle{F_q},G_p\rangle-\langle{F_p},G_q\rangle.
\]

The functions def\/ined by ($m=0,1,2,\ldots$)
\[
F_m=-\frac{1}{2}\langle\Lambda^{m+1}q,p\rangle-\frac{1}{2}
\sum\limits_{i+j=m}\left|
\begin{array}{cc}
\langle\Lambda^{i}q,q\rangle&\langle\Lambda^{i}q,p\rangle
\vspace{1mm}\\
\langle\Lambda^{j}p,q\rangle&\langle\Lambda^{j}p,p\rangle \end{array}
\right|
\]
are in involution in pairs (see, [9]).

Consider the Moser constraint on the tangent bundle
\[
TS^{N-1}=\left\{(p,q)\in {{\bf R}^{2N}}|F=\langle{q},p\rangle=0,
\ G=\frac{1}{2(\beta+1)}(\langle{p},p\rangle
-(-1)^{\beta}\langle{q},q\rangle^{\beta})=0\right\}.
\]
Through direct calculations we have
\[
(F,F_m)=0, \qquad (F,G)=\langle{p},p\rangle,
\] 
\[
(F_{m},G)=-\frac{1}{2(\beta+1)}\left(\langle\Lambda^{m+1}p,p\rangle+
(-1)^{\beta}\beta\langle{q},q\rangle^{\beta-1}
\langle\Lambda^{m+1}q,q\rangle\right).
\]
Thus the Lagrangian multipliers are
\[
\mu_{m}=\frac{(F_{m},G)}{(F,G)}=-\frac{1}{(\beta+1)}
\left(\frac{\langle\Lambda^{m+1}p,p\rangle}{\langle{p},p\rangle}
+(-1)^{\beta}\beta\frac{\langle{q},q\rangle^{\beta-1}}
{\langle{p},p\rangle}\langle\Lambda^{m+1}q,q\rangle\right).
\]
Since $F=0$ on the tangent bundle $TS^{N-1}$, the restriction of the
canonical equation of $H^{\ast}=F_0-\mu_{0}F$ on $TS^{N-1}$ is
\[
\left\{ \begin{array}{l} q_{x}=F_{0,p}-\mu_{0}F_{p}|_{TS^{N-1}},
\vspace{2mm}\\
p_{x}=-F_{0,q}+\mu_{0}F_{q}|_{TS^{N-1}} 
\end{array} \right.  
\]
which is exactly the Neumann system (4.7).

\medskip

\noindent
{\bf Theorem 5.1.} {\it 
The Neumann system $(4.7)$ $(TS^{N-1},dp\wedge{dq}|_{TS^{N-1}},
H^{\ast}=F_0-\mu_{0}F)$ is completely integrable in Liouville sense.}

\medskip

\noindent
{\bf Proof.} Let $F_{m}^{\ast}=F_{m}-\mu_{m}F$, $m=1,\ldots,N-1$, then it is
easy to verify $(F_{k}^{\ast},F_{l}^{\ast})=0$ on $TS^{N-1}$. Hence
$\{F_{m}^{\ast}\}$ is an involutive system.

\setcounter{equation}{0}
\section{Integrability of the Bargmann system}  
Let
\begin{equation}
\Gamma_{k}=\sum\limits_{\mbox{\scriptsize$\begin{array}{c}j=1\\
j\not=k\end{array}$}}^{N}
\frac{B^{2}_{kj}}{\lambda_k-\lambda_j},
\end{equation}
where $B_{kj}=p_kq_j-p_jq_k$, we have (see Refs.~[9, 10])

\medskip

\noindent
{\bf Lemma 6.1.} 
\begin{equation}
\left (\langle{q},p\rangle,p_l^2\right)=2p_l^2,
\qquad 
\left(\langle{q},p\rangle,q_l^2\right)=-2q_l^2,
\end{equation}
\begin{equation}
\ba{l}
\ds \left(p_k^2,\Gamma_l\right)=\frac{-4B_{lk}}{\lambda_l-\lambda_k}p_{k}p_{l},
\qquad 
\left(q_k^2,\Gamma_l\right)=
\frac{-4B_{lk}}{\lambda_l-\lambda_k}q_{k}q_{l},
\vspace{3mm}\\
\ds  \left(q_{k}p_{k},\Gamma_l\right)=\frac{-2B_{lk}}{\lambda_l-\lambda_k}
(p_{k}q_{l}+q_{k}p_{l}).
\ea
\end{equation}

\medskip

\noindent
{\bf Lemma 6.2.}
\begin{equation} 
(\Gamma_k,\Gamma_l)=(\langle{q},p\rangle,\Gamma_l)
=(\langle{q},p\rangle,q_lp_l)=0, 
\end{equation}
\begin{equation}
\left(p_k^2,p_l^2\right)=\left(q_k^2,q_l^2\right)=(q_kp_k,q_lp_l)=0,
\end{equation} 
\begin{equation}
\left(q_kp_k,p_l^2\right)=
2p_kp_l\delta_{kl}, \qquad 
\left(q_k^2,p_l^2\right)=4q_kp_l\delta_{kl},
\qquad \left(q_k^2,p_lq_l\right)=2q_kq_l\delta_{kl}.  
\end{equation}

\medskip

\noindent
{\bf Proposition 6.1.} {\it Let
\[
E_k=\frac{1}{2}\langle{q},p\rangle^{\frac{1}{\beta+1}}p_{k}^{2}
-\frac{1}{2}\langle{q},p\rangle^{\frac{\beta}{\beta+1}}q_k^2
-\frac{1}{2}\lambda_k{q}_k{p}_k-\frac{1}{2}\Gamma_{k}, 
\]
the
$E_1,\ldots,E_N$ constitute an $N$-involutive system.}

\medskip

\noindent
{\bf Proof.} Obviously $(E_k,E_l)=0$ for $k=l$. Suppose $k\not=l$, in
virtue of (6.4)--(6.6) and the property of Poisson bracket in $({\bf
R}^{2N},dp\wedge{dq})$, we have
\[
\ba{l}
\ds 4(E_k,E_l)=\frac{1}{\beta+1}p_k^2\langle{q},
p\rangle^{\frac{1-\beta} {\beta+1}}
\left(\langle{q},p\rangle,p_l^2\right) +\frac{1}{\beta+1}p_l^2
\langle{q},p\rangle^{\frac{1-\beta}{\beta+1}}
\left(p_k^2,\langle{q},p\rangle\right)
\vspace{3mm}\\
\ds \phantom{4(E_k,E_l)=} -\frac{1}{\beta+1}p_k^2\left(\langle{q},p\rangle,q_l^2\right)
-\frac{\beta}{\beta+1}q_l^2\left(p_k^2,\langle{q},p\rangle\right)
-\langle{q},p\rangle^{\frac{1}{\beta+1}}
\left(p_k^2,\Gamma_l\right)
\vspace{3mm}\\
\ds \phantom{4(E_k,E_l)=} -\langle{q},p\rangle^{\frac{1}{\beta+1}}
\left(\Gamma_k,p_l^2\right)
 -\frac{\beta}{\beta+1}q_k^2\left(\langle{q},p\rangle,p_l^2\right)
-\frac{1}{\beta+1}p_l^2\left(q_k^2,\langle{q},p\rangle\right)
\vspace{3mm}\\
\ds \phantom{4(E_k,E_l)=}+\frac{\beta}{\beta+1}q_k^2\langle{q},p\rangle
^{\frac{\beta-1}{\beta+1}}\left(\langle{q},p\rangle,q_l^2\right)
+\frac{\beta}{\beta+1}q_l^2\langle{q},p\rangle
^{\frac{\beta-1}{\beta+1}}\left(q_k^2,\langle{q},p\rangle\right)
\vspace{3mm}\\
\ds \phantom{4(E_k,E_l)=}  +\langle{q},p\rangle^{\frac{\beta}{\beta+1}}
\left(q_k^2,\Gamma_l\right)+\langle{q},p\rangle^{\frac{\beta}{\beta+1}}
\left(\Gamma_k,q_l^2\right)
+\lambda_k(q_kp_k,\Gamma_l)+\lambda_l(\Gamma_k,q_lp_l). 
\ea
\]
Substituting (6.2) and (6.3) into the above equation yields
$(E_k,E_l)=0$.

Consider a bilinear function $Q_z(\xi,\eta)$ on ${\bf R}^N$:
\[
Q_z(\xi,\eta)=\langle(z-\Lambda)^{-1}\xi,\eta\rangle
=\sum\limits_{k=1}^{N}\frac{\xi_k\eta_k}{z-\lambda_k}
=\sum\limits_{m=0}^{\infty}z^{-m-1}
\langle\Lambda^{m}\xi,\eta\rangle.
\]
The generating function of $\Gamma_k$ is (see, [9, 10])
\[
\left| \begin{array}{cc} Q_z(q,q)&Q_z(q,p)\vspace{1mm}\\ 
Q_z(p,q)&Q_z(p,p)
\end{array}
\right|=\sum\limits_{k=1}^{N}\frac{\Gamma_k}{z-\lambda_k}.
\]
Hence the generating function of $E_k$ is
\begin{equation}
\ba{l}
\ds \frac{1}{2}\langle{q},p\rangle^{\frac{1}{\beta+1}}Q_z(p,p)
-\frac{1}{2}\langle{q},p\rangle^{\frac{\beta}{\beta+1}}Q_z(q,q)
-\frac{1}{2}Q_z(\Lambda{q},p) 
\vspace{3mm}\\
\ds  \qquad \qquad -\frac{1}{2}\left| \begin{array}{cc}
Q_z(q,q)&Q_z(q,p)\vspace{1mm}\\ Q_z(p,q)&Q_z(p,p) \end{array}
\right|=\sum\limits_{k=1}^{N}\frac{E_k}{z-\lambda_k}.
\ea
\end{equation}
Substituting the Laurent expansion of $Q_z$ and
\[
(z-\lambda_k)^{-1}=\sum\limits_{m=0}^{\infty}z^{-m-1}\lambda_k^m
\]
in to both sides of (6.7) respectively, we have

\medskip

\noindent
{\bf Proposition 6.2.} {\it Let
\[
F_m=\sum\limits_{k=1}^{N}\lambda_k^mE_k, \qquad m=0,1,2,\ldots
\]
then
\[
F_0=\frac{1}{2}\langle{q},p\rangle^{\frac{1}{\beta+1}}\langle{p},p\rangle
-\frac{1}{2}\langle{q},p\rangle^{\frac{\beta}{\beta+1}}
\langle{q},q\rangle-\frac{1}{2}\langle\Lambda{q},p\rangle,
\]
\[
\ba{l}
\ds F_m=\frac{1}{2}\langle{q},p\rangle^{\frac{1}{\beta+1}}
\langle\Lambda^{m}p,p\rangle-\frac{1}{2}\langle{q},p\rangle
^{\frac{\beta}{\beta+1}}\langle\Lambda^{m}q,q\rangle
\vspace{3mm}\\
\ds \phantom{F_m=} -\frac{1}{2}\langle\Lambda^{m+1}q,p\rangle
-\frac{1}{2}\sum\limits_{j=1}^{m} \left| \begin{array}{cc}
\langle\Lambda^{j-1}q,q\rangle&\langle\Lambda^{j-1}q,p\rangle
\vspace{1mm}\\
\langle\Lambda^{m-j}p,q\rangle&\langle\Lambda^{m-j}p,p\rangle
\end{array}
\right|.
\ea
\]
Moreover, $(F_k,F_l)=0$.}

\medskip

Hence we arrive at the following theorem.

\medskip

\noindent
{\bf Theorem 6.1.}
{\it The Bargmann system defined by $(4.9)$ is completely integrable in
Liouville sense in the symplectic manifold $({\bf
R}^{2N},dp\wedge{dq})$.}

\subsection*{Acknowledgement}

I am very grateful to Professor Cao Cewen for his guidance. This
project is supported by the Natural Science Fundation of China.

\label{jiang-lp}

\end{document}